# Beyond the Stalemate: Conscious MInd-Body - Quantum Mechanics - Free Will - Possible Panpsychism - Possible Interpretation of Quantum Enigma

Stuart Kauffman Oct 20, 2014

**Introduction**

I wish to discuss a large, interwoven set of topics pointed at in the title above. Much of what I say is highly speculative, some is testable, some is, at present, surely not. It is, I hope, useful, to set these ideas forth for our consideration. What I shall say assumes quantum measurement is real, and that Bohm's interpretation of Quantum Mechanics is not true.

The Stalemate: In our contemporary neurobiology and much of the philosophy of mind post Descartes we are classical physics machines and either mindless, or mind is at best epiphenomenal and can have no consequences for the physical world. The first main point of this paper is that we are not forced to this conclusion, but must give up total reliance on classical physics. Please note that this avXiv version further develops ideas on line and in press under the same title, with Sean O'Nuallain as editor.

**The Causal Closure of Classical Physics Is The Source of the Stalemate**

We all know Newton, his three laws of motion, universal gravitation, and invention of differential and integral calculus. Given seven billiard balls rolling on a billiard table, we might ask Newton what will happen to the balls. "Write down the initial conditions of position and momenta of the balls, the boundary conditions of the edges of the table, and the forces between the balls, and the balls and the edge of the table using my three laws of motion in differential equation form. Then, to find out what will happen to the balls in the future (or past, my laws are time reversible), integrate my differential equations to obtain the trajectories of the balls (for all time in the absence of friction)". But, I note, integration is deduction of the consequences of Newton's differential equations for the trajectories of the balls, and deduction is "entailment". "All men are mortal. Socrates is a man, therefore Socrates is a mortal." is a syllogism whose conclusion is logically entailed by the truth, if so, of the premises. So too the trajectories are entailed by integration of Newtons differential equation.

But this entailment sets up the Stalemate. If the brain is a classical physics system, then the present state of the classical physics brain is entirely sufficient to determine the next state of the brain. But then, there is NOTHING for mind to do, and NO WAY for mind to do it! It would be like asking mind to alter the trajectories of the balls on the billiard table.

Thus, if mind somehow is present in a classical physics setting, it can have NO consequences at all for the classical physics world. At best, the mind can be merely epiphenomenal. (We might wonder if mind exists and is merely epiphenomenal, and if mind with brain evolved, what selective advantage could it have had?).

The culprit is the causal closure of classical physics with, as Aristotle said, no Prime mover. The Stalemate arises because we want mind to act causally on brain, but it cannot because all the classical physics causes are already in the laws of the billiard ball classical physics neuronal system and attendant classical physics further variables including classical physics noise.

**Quantum Mechanics Provides Two Ways to Break The Causal Closure of Classical Physics and Have ACAUSAL Consequences for the "Classical" Brain**

I begin with a familiar outline of Quantum Mechanics, with the caveat that I am not a physicist. 1) We all know the two slit experiment and the resulting interference pattern of spots on the developed film emulsion beyond the two open slits. 2) We know the Schrodinger linear wave equation, often set equal typically to a classical potential V. The equation has no energy term, so what is "waving" cannot be matter or energy. No one knows what is "waving". I will propose below that what are waving are "possibilities", (1), or, with Heisenberg, Potentialities,(2). 3) We know the Born rule: square the amplitude of each wave, say spin up or spin down, in superposition, and that is the probability that upon measurement that outcome will be found. We know there are 16 interpretations of Quantum Mechanics, in which measurement is real in some and not others. As noted above I assume measurement is real. 4) Finally we all know the astonishing confirmation of Non-locality for entangled quantum variables.

**Res potentia and Res extensa, Linked, hence United, by Measurement**

I may be proposing a 17th interpretation of quantum mechanics, rather similar in some aspects to Heisenberg's Potentia,(2), but on different grounds. I assume quantum measurement can be real in general, and in particular, for a subset of entangled variables, and is ontologically indeterminate. On this view Bohm is not correct, nor is the Multiple World interpretation. Of course there are many interpretations of Quantum Mechanics. I here explore a new one in some respects. I begin with Feynman's formulation of quantum mechanics as a sum over all possible simultaneous histories,(3). This is accepted as an equivalent formulation of quantum mechanics by most physicists. On this formulation, one must say that a SINGLE photon on its way through the two slits to the film emulsion simultaneously does and does NOT pass through the left slit. But this statement breaks Aristotle's Law of the Excluded Middle, where "A and Not A" is a contradiction. Thus, on Feynman's formulation, quantum coherent behavior does NOT obey the law of the excluded middle. After quantum measurement, the result does obey the law of the excluded middle. The same holds for simple superpositions of single or entangled particles. For example, a single independent electron, in a superposition state alone, or if entangled, is ontologically spin up with 50% probability by the Born rule and 50% down, upon measurement, then is found to be either spin up, or spin down, not both simultaneously. So measurement, if real, does take Quantum Mechanics from something that does not obey the law of the excluded middle to something that does obey the law of the excluded middle.

Philosopher C. S. Pierce noted that Actuals and Probables DO obey the law of the Excluded Middle, but Possibles do not. Thus, "The photon possibly did and possibly did not simultaneously go through the left slit" is NOT a contradiction.

I now wish to propose a new dualism, but not a substance dualism: Res potentia - ontologically real Possibles that do not obey the Law of the Excluded Middle, and Res extensa - ontologically real Actuals that do obey the Law of the Excluded Middle, linked by measurement, (1). The Schrodinger equation has no term for energy, thus whatever is "waving" is not matter or energy. Because Possibles are not a substance, consistent with the fact that the Schrodinger equation has no energy term, this dualism is not a substance dualism. "Possibles" are not far from Heisenberg's "Potentia",(2), indeed may be identical. If so, what is waving in the Schrodinger equation are possibilities. "Where" are possiblities?" Where is the possibility you will fall in the next 10 minutes? It seems that possibilities cannot be located in space, but perhaps in time, consistent with Special Relativity.

I note next that if we accept Res potentia and Res extensa united by measurement, there can be no deductive mechanism for measurement. The "X is possible" of Res potentia does not deductively entail the "X is actual" of Res extensa. Physicists like equations. Here the logic is very simple, but correct. This is an hypothesis easily disproved were we to discover a deductive mechanism for measurement, not found since 1927, but a valid test of the Res potentia hypothesis only assuming measurement is real.

A puzzle in quantum mechanics is why superpositions, as such, are never directly measured. If we try the hypothesis of Res potentia and Res extensa, then measurement converts possibles to actuals, the former not obeying the law of the excluded middle, the latter obeying that law. We cannot measure possibles. A bit of care. If we measure the energy of an electron in an atom, we infer that its momentum, a complementary aspect, is dispersed. But we have not measured the momentum itself.

**Res potentia, Instantaneous Changes in the Wave Function upon Measurement, and Non-Locality**

We have a hard time thinking about non-locality where measurement of one entangled particle can instantaneously alter the outcome to be found upon measurement of the other entangled particle even if spatially far away, ie non-local causally. In addition a puzzling aspect of quantum mechanics is that when a measurement happens, the wave function changes "instantaneously", even for independent particles. How can these changes be "instantenous" in the mathematics of quantum mechanics. Neither for the single nor entangled particles, can this be "causal". For non-locality, workers suggest "influences" or "information" that travels faster than the speed of light.

I now give an informal every day example. Suppose you wish to go to the store on J street tomorrow to meet Bill and buy orange juice. Today you go by the store just as a sign goes up on the store, "Store closed immediately!". What just happened to the possibility that you could, tomorrow, go to the store, meet Bill and buy orange juice? The   Possibility just vanished! More the possibility vanished when a Actual changed, ie the store closed "immediately". Thus, let's consider that a change in an actual immediately and entirely ACAUSALLY, changes what is now possible.

Note that, in fact, we live our lives believing that the above really is true of our world.

Another example which requires no implication about consciousness concerns "exaptations" in biological evolution. For example, swim bladders evolved from the lungs of lung fish and provide a new function, neutral buoyancy in the water column. But once the swim bladder exists as an Actual, it thereby creates, or "enables" new possibilities in the evolution of the biosphere. For example a worm or or bacterial species could evolve to live in the swim bladder once the bladder exists in the evolution of the biosphere, but not before. A change of actuals, here the swim bladder, alters the possible pathways of biological evolution. The same is true for technological evolution.

In both the J street store case, and the swim bladder, a changing actual changes what is now possible in further evolution, can do so instantaneously, and the "becoming possible" is acausal!

Now consider Res potentia and Res extensa, linked by measurement, where measurement is assumed real and ontologically indeterminate. When the electron is measured and found to be spin down, it is now an Actual that obeys the law of the excluded middle, hence is a new actual that creates new possibles instantaneously and acausally. Thus on this view, a new actual alters what is possible, so instantaneously and acausally changes the wave function, including for a single variable now measured, which flowers new amplitudes, ignoring the Zeno effect, or for space-like separated entangled particles. Thus, the hypothesis of ontologically real possibles offers a new explanation of what is waving in the Schrodinger equation, possibilities, and how they can change instaneously and non-locally, when an actual changes.

**The "Poised Realm" Hovering Reversibly Between Quantum and "Classical" Worlds**

Gabor Vattay, Samuli Niiranen and I have recently proposed, or perhaps discovered, a new "Poised Realm", (4,5), in which the total system can hover reversibly between quantum coherent and "classical" worlds, with the known debates about what the classical world may be. The Poised Realm is captured by an X Y coordinate system. At the origin on the Y axis, the system is quantum coherent. As the system moves up the Y axis, the system undergoes increasing decoherence as an open quantum system losing phase information to the environment and approaches "classicality" infinitely closely for all practical purposes, FAPP, (6). (I note that the decoherence program no longer seems to think it can account for quantum measurement, again if measurement is real, (7). What is new on the Y axis, is that REcoherence can occur and move the system back down the Y axis from "Classical" FAPP to quantum coherent. The possibility of recoherence is assured by a theorem by Peter Shor, (7), now in use for quantum error correction in decohering qubits with the input of "information". More recently, quantum biology at body temperature is firmly established in the long lived quantum coherence of light harvesting molecules. Even more recently, experimental evidence supports recoherence induced by phonons in the light harvesting complex that induce recoherence in the electron involved, (8). The Y axis seems real in theory and practice.

The X axis is: Order, Criticality and Chaos, going out from the origin. In classical physics, this arises as a set of Hamiltonians are tuned from a conservative oscillation like a pendulum, where neighboring orbits are parallel and the Lyapunov exponent is thus 0, to a critical point out the X axis where the Lyapunov exponent undergoes a second order phase transition to become slightly positive at Criticality, then more positive in the Chaotic parts of the X axis. In the quantum coherent world, criticality corresponds to the metal - insulator transition between extended and localized wave functions. The X axis is also real. Vattay and colleagues, (9),have now measured the absorption spectra of hundreds of different organic molecules. From each molecule one constructs a histogram of the number of small energy intervals between absorption bands, and larger and larger energy intervals. The ordered regime corresponds to a well known exponential decay, from may short intervals to a few long intervals. The Chaotic regime is a unimodal distribution. The critical regime is a unimodal distribution whose single peak is shifted toward shorter wavelengths than the chaotic regime, (5). Almost half the molecules examined are ordered, and, astonishingly, almost half are critical, a single point on the X axis. A few are chaotic. So the X axis is real. Why half are critical is a new mystery. Thus, the X axis is real. Since both the X and Y axis are real, the Poised Realm is real.

New physics arises in the Poised Realm, in part because decoherence is dissipative, so the Schrodinger equation does not propagate time reversibly and unitarily, but temporal behavior can be followed using Density Matrix methods. Experimental evidence for new physics includes the fact that in coherent systems, jumps between quantum states are Poisson distributed in time, yielding the familiar exponential half life. In the presence of decoherence, the jumps are no longer Poisson in time, confirmed experimentally, and sometimes called the AntiZeno Effect,(10). The full portent of the Poised Realm may be very large. I will propose below that it plays a major role in the Mind - Body problem.

**Two Ways the Poised Realm Allows ACAUSAL Consequences for the "Classical World"**

First, if we accept deoherence to classicality FAPP, then decoherence is entirely an acausal loss of phase information from the open quantum system to the universe. So a quantum "mind" can have acausal consequences for the classical "meat" of the brain, breaking the Stalemate of the Causal Closure of Classical Physics. But by recoherence, the total system, quantum, poised realm and "classical" can have repeated acausal consequences for the "classical meat of the brain". Here I take "mind" to include quantum coherence, measurement, and the Poised Realm. We will see below that a quantum description does not yet suffice to get us to "consciousness" or experiential terms including consciousness and responsible free will.

The second way that the above system can have acausal consequences for the "classical' meat of the brain is quantum measurement which can acausally restore coherence in a decohering quantum variable, or, again if measurement is real, alter the classical world by leaving a record, as the spot in the two slit experiment does.

In short a Mind - Body system that is quantum coherent, Poised Realm, and "classical", can escape the Stalemate. Mind can, in principle, be real in the world and effect its becoming, and not merely epiphenomenal. This alone suggests taking any such ideas and considering them. We have been frozen with the Stalemate for 350 years.

However, the discussion above is void of experiential terms such as consciousness, qualia, and responsible free will, topics I take up below.

**Beginning Implications for Neurobiology**

With Penrose, (11) and with Hammeroffs' Objective Reduction for measurement accompanied with flashes of consciousness at measurement but no "doing", hence an epiphenomenal panpsychism, and Stapp,(12) and measurement with the Zeno effect, but for overlapping reasons, I want to propose that conscious experience, qualia, is associated with quantum measurement. I do so in part based on the fine discussion of the Quantum Enigma, (12), discussed with hesitation below. The first important issue is that we never consciously experience the quantum coherent state, rather we SEE the results of measurement. This suggests that conscious experience is not associated with quantum coherence itself, but with a "yes" outcome to quantum measurement. I discuss below the potentially testable relation between conscious human experience and "measurement" for a rod can detect absorption of a single photon, but conscious experience also needs higher brain centers. We also want "instruments" to measure without human consciousness. I try to address this below as well.

The hypothesis that conscious experience is associated with measurement is also experimentally testable genetically, (1). Fruit flies, and other animals, can be anesthetized by ether. Take a wild type population of fruit flies, and over generations, select a "mutant" subpopulation that can be anesthetized by shorter and lower doses of ether, perhaps until none is needed. Sequence the DNA of the selected "mutant" and wild type populations of flies, detect mutant genes, hence proteins, if they exist, in the selected population, and test if those proteins fail to carry out some quantum measurement that the wild type proteins do "in vivo". I do not know how to pick out what quantum measurements might be relevant, but clearly this is a start of an experimental test.

**Connections to Neurobiology**

The anatomical location of such "mutant" proteins can be established. Suppose the relevant proteins were located in synapses and part of the post cleft neurotransmitter protein receptor complex. Then one can imagine that quantum behaviors altering the receptor protein(s) could affect adjacent dendritic transmembrane potentials, the subsequent potentials transmitted to neural cell bodies and summing, or not, at the axon hillock to trigger action potentials propagating down the axons. In principle, this would be a start of tying quantum behavior to all of classical neurobiology including our highly evolved sensory and motor systems and myriad brain structures. But the proposal that conscious experience is associated with measurement "turns the brain on its head". The business end of the mind brain system related to conscious qualia are the relevant quantum variables in the brain. However, I need only claim that measurement is a necessary, not sufficient condition for consciousness. As noted recent experiments with stimuli that can shift from "conscious of" to "not conscious of", suggest that higher brain areas are also necessary for "conscious of".

**Entanglement and the Binding Problem**

In the Astonishing Hypothesis, Francis Crick describes the binding problem, (14j). We are to suppose that a yellow triangle and blue square are being observed. If "yellow", "blue", "triangle" and "square" are processed in anatomically disconnected areas of the brain, a claim I will accept, then how do "yellow" and "triangle" become bound into the experience of "yellow triangle" and how do "blue" and "square" become bound into the experience of "blue square". One solution is a 40 Hertz oscillations in the brain and if "yellow" and "triangle" occur at one phase of the oscillation and "blue and "square" at a different phase, then binding occurs to yield "yellow triangle" and "blue square". This might work. My problem is that we seem to bind indefinitely many, here visual, qualities or features and fitting them all into "discriminably" different phases of the oscillation seems problematic, but conceivable.

I want to try entanglement to solve the binding problem, (1). There is now weak evidence for entanglement of a number of quantum variables, including photons, in cells and the brain, (15). Suppose that some set of N quantum variables in the brain can become entangled and are in anatomically disconnected areas of the brain. Then it is known that upon measurement, their outcomes are correlated and more correlated as the number of entangled variables increases. Thus the qualia upon measurement may be highly correlated into a new "whole". This could mediate "binding" in anatomically disconnected areas of the brain in ways that also respect neuroanatomical connectivity.

The binding problem is a subset of the Unity of Consciousness issue, for example, the experienced perceived whole visual field. We might suppose that this unity of consciousness is mediated by entanglement and co-measurement and an enormously rich evolved neural system. Such proposed entanglement is, in principle, testable. Three further issues arise, for we experience a "unity of consciousness". 1) But with shifting attention we seem to experience different unities of consciousness. Then we might imagine that the shifting attention and resulting shifting unity of consciousness reflects in part different subsets of quantum variables becoming entangled and measured to yield shifting patterns of conscious experience, presumably tied to known neural processes related to attention. 2) Entanglement solves the "combination" problem of W. James, who considered atoms of consciousness, but then noted that combinations of consciousness atoms seemed to yield new "wholes". He noted that a house made of bricks, from the viewpoint of the bricks is just a bunch of bricks, not a new "whole", but only a new "whole",  say a house, to an outside observer. But there is no outside "observer". James never solved the combination problem. But with entanglement, the N entangled variables are no longer independent at all, so can give rise to an experience which is a new whole, perhaps solving the combination problem. The new whole may also solve the famous Frame Problem of computer science. 3) In order for entanglement to work and to shift, there must be means to entangle different "desired" sets of quantum variables. I give one conceivable mechanism: decoherence broadens absorption bands, recoherence narrows them. Thus imaging entangling a "desired" set of decoherent quantum variables in a wavelength about 300K, the temperature of the brain so able to transverse the brain, but chosen so that wavelength will be absorbed only in the broadened absorption region of decoherent quantum variables. Then by decoherence and recohrence, which variables become entangled can be altered.

**Possible Direct Tests of Quantum Mind**

If mind is partially quantum coherent or also Poised Realm, then by entanglement and non-locality, both telepathy and telekinesis are possible. Weak evidence and somewhat improved evidence supports both, (16, 17). Indeed, there are reports of many computers around the world, generating random numbers both with classical random number generators, and by quantum randomness, tested for "erratic" behavior when major public events happen, such as the death of Nelson Mandela. It is claimed that the resulting deviations are statistically significant. I cannot attest to this, but this kind of experiment has the strong virtue that the outcomes, if statistically significant outputs of computers, are objective data. By contrast, claimed telepathic experiences are harder to verify. Such telekinesis experiments are quite cheap. If confirmed, the obvious explanation is non-locality, hence a quantum role in the mind-body system.  It seems very worth pursuing such experiments, for the causal closure of classical physics can only yield an epiphenomenal mind, while a quantum mind can have consequences for the "classical" body and world, so be a mind beyond merely epiphenomenal. Decohrence may render such phenomena evanescent, a topic perhaps to be examined.

**The Strong Free Will Theorem and The Subjective Pole**

Nothing in the above discussion involves "experience" in any way. I have at best afforded a way to achieve a "mind body" system in which  quantum and Poised Realm mind really can alter the "classical" world by acausal consequences for brain, so mind need not be merely epiphenomenal. That is a lot, if true.

Since Descartes's Res cogitans, in his substance dualism with Res extensa,

failed and Res extensa and Newton won with nothing for mind to do and no way for mind to do it given the triumph of classical physics and its causal closure, the subjective pole of experience has been distanced by the "hard sciences".

But a recent theorem by Conway and Kochen, two mathematicians, The Strong Free Will Theorem, (18), offers its own hope for the "experiential pole". The theorem states that if two physicist have independent free will with measurements of two entangled spin 1 particles, physicist A measuring the squared spin component in three orthogonal directions of spin particle "a", and physicist B happens with free will to measure spin particle "b" in one of the three directions measured in "as", with free will, then: 1) Nothing in the past of the universe determines the outcome of measurements. 2) There can be no mechanism for measurement (independently supported by Res potentia and Res extensa linked by real measurement. 3) Then the stunning: The two entangled electrons free will jointly free will DECIDE! This is the only theorem I know in physics that uses an experiential term, "Decides", but it depends upon measurement being real.

Assume for the moment the theorem holds. Then there is a further mystery. If 1000 electrons are identically prepared and independent, to be measured spin up with 50% probability and spin down with 50% probability and all independent electrons are measured, in fact, about 50% will be spin up and 50% will be spin down with a Gaussian distribution about this mean. But if the electrons are independent and each has "free will" and "decides", how can it turn out that in fact about 50% wind up spin up and about 50% wind up spin down? The easiest thought is that each electron has a "preference" to choose with 50% spin up and choose with 50% spin down. If so, free will "decides" comes with "preferences" and hence a "will" that is "responsible".

But the Strong Free Will Theorem is circular. It assumes the two physicists have free will then proves the electron does. But why should we assume the physicists has free will? I now want to try to break this circularity on an ontological, not yet experiential, basis.

Here are the two standard "responsible free will" issues. If classical physics holds we have no free will at all. More our normal sense of free will is that we could have done otherwise. But this cannot arise in classical physics, for "could have done otherwise" requires that statements that are conditionals contrary to fact can sometimes be true. In classical physics, given initial and boundary conditions, as in the billiard balls on the table, all that happens is what ACTUALLY happens. The behavior could not have been otherwise, hence the presence could not have been different. If we try the move: If the initial conditions, or boundary conditions were different, what actually happens could have been different, it seems not to work. Classical physics assumes the boundary condition and initial conditions are given "somehow". But how could the initial conditions "have been different" so the present could have been different? Within classical physics, there is no answer.

If we try to use quantum indeterminism to achieve an ontologically free will, it is merely random. But a responsible free will? It is claimed, NO. So a random quantum event occurs in my brain, I kill the little old lady, but I am not "responsible", the quantum event was random. So even if measurement is real, and ontologically indeterminate, so underlies a "free will", that will cannot be responsible. Before I try to rescue the standard worry about quantum random indeterminism in the free will discussion below, I must make a central point. Free will seems to require the truth, sometimes, of counterfactual statements such as "I could have done otherwise" eg turn left not turn right while driving. But such statements demand that the present moment could have been different. This is not possible in a deterministic world view, for the becoming of the universe since its inception can only have happened one way, hence the present, if time is real, cannot have been different. (It is not useful to say the present could have been different in a deterministic world with different initial conditions, for these different initial conditions arise without explanation.) But IF quantum measurement is real and ontologically indeterminate, as it is on most interpretations of quantum mechanics, including Res potentia and Res extensia linked by measurement, then the present could have been different: The electron could have been measured as spin up, or as spin down, so the present could have been either of these. I think real, ontologically indeterminate measurement is a condition consistent with the possibility of free will.

But quantum events are random. Can one find an ontological, not yet experiential, response to the "random" aspect of the objection to quantum measurement precluding a "responsible free will"? I think probably "Yes". My discussion now is to consider a "set" of quantum measurements", show this fails if the N quantum systems are independent, but can succeed if the N quantum systems are entangled, hence the set of measurements outcomes are not independent.

The situation is not helped if we consider N = 50 independently prepared electrons, all say 50% spin up and 50% spin down, and measure them all. We will get an INDEPENDENT set of 50 random up/down outcomes. So again, the set as a whole, whose outcomes are random and independent, cannot yield an ontological basis, not yet experiential, for a responsible free will. The set is as random, as N INDEPENDENT measurements, as a single measurement.

Now consider N entangled particles, say electrons, again spin up and spin down. It is a standard theorem of Quantum Mechanics for entangled particles that measurement of the first particle, whatever the outcome, up or down, typically

CHANGES the probabilities for the outcome of the next measured electron and so on for all N electrons. This arises because the first measurement alters the density matrices used to propagate the Schrodinger equation for the remaining N - 1 electrons. Thus, each successive measurement of the next among the N entangled particles, alters, by the Born rule, the probabilities of the next measurement. So the "set"of N measurement outcomes are NOT independent. Indeed this non-independence is now one new approach to quantum computing, (19).

I now argue that this non-independence provides an ontological basis, not yet experiential basis, for a responsible free will. Below I use known facts about quantum measurement fitting the Born rule to argue that, if the choices are free, they reflect something like a "preference", hence are "protoresponsible" as well as free. First, note that in a limiting case, the amplitudes could be altered such that the probabilities for spin up or spin down went all the way to 100% or 0%.Thus the first measurement could have been 100% up, changing the second measurement to 100% down, changing the third measurement to 100% down for all N measurements. This limit is central, for in this case, the outcomes of real, ontologically indeterminate, hence "Free" measurements is entirely NON - RANDOM. Hence the claim that ontologically real and, "free because they are indeterminate, measurements, are merely random is here fully defeated. I take it that this suffices to render the standard objection to quantum measurement, its randomness, no longer an obstacle to an ontological basis for a responsible free will. As hinted above and discussed just below, the fact that the "free choices" obey the Born's rule strongly suggests that the deciding particle(s) have a "preference", hence the choices are "protoresponsible". In the 100% 0% limit we here consider the choices of the N entangled particles are free and, I argue, responsible, but only a single such choice is made for each of the N measurements, so for the set as a whole. But this 100% and/or 0% for each of the N entangled measurements is not what we want either, in general. We want for the responsible decision or choice of outcomes to include more than one choice, each made responsibly. Thus, in general, let the probabilities of the N entangled particles change, but typically remain less than 100% or greater than 0%. Then the outcome of each measurement is ontologically indeterminate, but the "set" of N entangled measurements is no longer independent. These Born probabilities, none equal to 100% or 0%, constitute what I want to call "enabling constraints" on the deciding electron to "decide" within the constraints set by the Born rule. I believe this is enough to remove the "randomness" objection to an ontologically indeterminate, quantum based free will. The measurement outcomes of the N entangled particles are not independent whatever the Born probabilities may be and change to over the set of N measurements.

The above is not yet quite enough, as I now explicitly try to state, to entirely decircularize the strong free will theorem which also assumes the two physicists can independently free will measure the two entangled spin 1 particles to be measured, particle "a" measured by physicist A in three orthogonal directions, and "b" , free will measured independently by physicist B, in some direction "w", which may happen to be one of the three directions in which spin particle "a" was measured. It is essential that the Strong Free Will Theorem does NOT require that the two physicists independently decide "responsibly". The theorem requires only that the two physicists can INDEPENDENTLY FREE WILLED measure the pair of entangled spin 1 particles as just described. But the theorem does NOT require that the two independent physicists "responsibly" so decide. It suffices that at least sometimes they independently do free willed measure the first particle in three orthogonal axes, and measure the second in an axis, "w" that happens to coincide with one of the axes used to measure the first particle. But for this, it is entirely sufficient if for the first physicist, a single random quantum event, or an entirely 100%, 0% non-random entangled determinate set of N measurements yields that the first entangled spin 1 particle is measured in "some" set of three orthogonal axes, and similarly a quantum random or entirely 100%, 0% entangled non-random set of measurements in the second physicist yields that the second entangled spin 1 particle is measured in an axis "w" which may happen to coincide with one of the axes used to measure the first particle. In these cases, with NO "responsible free will", the Strong Free Will Theorem holds and nothing in the past of the universe determines the outcomes of the measurements, there is no mechanism for measurement and the entangled electrons "freely decide". No "responsibility" is yet needed on the part of the two independently free willed physicists.

The above leads us to take the Strong Free Will theorem seriously, and from it we are led to a "responsible free will" on the part of the electrons and, I now suggest in humans as well.

I now comment on the very concept of "responsible free will" for humans. I claim that almost all human free will actions occur in the context of enabling constraints that create an  "adjacent set of possible relevant actions".  "I want to go to the store to buy food", and the enabling constraints include all the ways I could get to the store, drive my car, walk, skate, parachute, hitch-hike, bike, and my learned capacities to drive, walk, skate, parachute, hitchhike and so forth. Unconstrained "free will", in which we "can freely do literally anything at all", seems to make no sense for acting humans. How and just what would we "responsibly choose to do" out of the unstateable set of "literally anything at all"?  The question seems ill-posed. We cannot "responsibly decide" among an unstateable set of "literally anything at all". We, at least sometimes, responsibly act for reasons, purposes, and intentions, of which we are sometimes consciously aware, given by context and capacities that constitute, in part, the enabling constraints. Sometimes we are not aware consciously of our reasons purposes and intentions. Shakespeare understood this range of conscious and unconscious reasons, motives, and purposes full well in his plays, which better describe our humanity than science. Thus, and this claim is essential: I claim I have fully decircularized the Strong Free Will theorem that allows an ontological basis for a responsible free will for the physicist and for the electron!

But are we forced to accept the experiential term "will" by the theorem. I fear not yet. No third person description can entail an experiential term like Will, either for the physicist or the electron. We have here two choices: Deny "will" or accept "will". Below I discuss testing whether human conscious observation can be sufficient for measurement. In principle, as noted, we can test this. As we will see, if we take the quantum enigma to be real, then we humans do have free will and could have done either of two experiments, observe the quantum variable, or instead, we could, contrary to fact, have done the experiment demonstrating by inference, the interference pattern. All physicists seem convinced that "They could have chosen to do the other experiment, contrary to fact", yet proceed with no attempt to account for this. But if physicists can have responsible free will and have chosen otherwise, contrary to fact, then the Strong Free Will Theorem demands that we accept Free Will for electrons. I can see no experiment to demonstrate this for electrons at present, although I can imagine experiments convincing us that human conscious experience can suffice for measurement, also a part of the quantum enigma. Note that Penrose and Hammeroffs' OrchR and moment of consciousness which yields a panpsychism that is merely epiphenomenal and cannot alter the universe. If we accept that electrons can measure one another, then either we accept that they too mediate measurement acausally by conscious observation, or imagine another means electrons measure, when human conscious can, we can test, suffice. Therefore,I will accept the Enigma all the way from particles to us, we and they, with the Strong Free Will Theorem, will be conscious and thereby acausally mediate measurement, and "act", so propose we accept Responsible free will for electrons. This will lead to a broad new formulation of quantum mechanics in terms a new triad: Actuals, Possibles and Mind - conscious observation acausally mediating measurement, and doing. Here new actuals create new possibles which are available via mind to be measured to create new actuals to create new possibles in a persistent becoming of the universe. I return to this below. We are not forced to accept this view, but parts are testable and leads to a vast panpsychism in which the universe persistently and in part "freely" within Born rules and perhaps beyond for us, becomes at many levels by the Triad above. This leaves the classical world to account for, with many proposals.

Since Descartes' Res cogitans, a substance dualism, lost to his Res extensa via Newton, we have entirely lost the "subjective pole" in physics. We have, I think, now, partially regained it, but in a striking way.

**Human Knowing Beyond Propositions and the Law of the Excluded Middle**

In an article, "No entailing laws, but enablement in the evolution of the biosphere" by Giuseppe Long, Mael Montevil, Stuart Kauffman, (20), we hope we prove that we cannot prestate the ever changing phase space of biological functionalities in evolution. As a consequence, we can write and integrate no entailing laws of motion for the becoming of the biosphere, a major negative result if correct. The result states that specific evolution of the biosphere is not even mathematizable. If so no final theory can entail the becoming of the entire universe of which the biosphere is a part. More we show that we cannot in general, prestate the relevant biological functionalities that arise during evolution in what we call an "adjacent possible". For example swim bladders unprestably evolved from the lungs of lung fish, (21). Given feathers, co-opted from thermoregulation for flight, flight based on feathers was in the new and unprestatable Adjacent Possible of the evolution of the biosphere once there were feathers as new actuals. In general, we do not know ahead of time the possibilities in the Adjacent Possible in biological or in, e.g. technological, economic, or scientific evolution. If we do not know what CAN happen, we cannot reason about it. Sufficient reason fails us. Yet we make our way, not knowing what can happen. How?

How? Language evolved from metaphoric language to propositional language. We all use metaphors to orient and act in the world. Art is metaphoric. Now, critically, metaphors are neither true nor false! This may well echo quantum coherent behavior being neither true nor false, ie not following Aristotle's law of the excluded middle. There is evidence that human concept usage has quantum, not classical logic, (22). Further, note that it seems clear that no prestated set of true/false propositions can exhaust the meaning of a metaphor.

Now consider the invention of propositional, true false, language. With it comes the possibility of logic and syllogism. "All men are mortal. Socrates is a man. Therefore Socrates is a mortal". From propositions comes also the later emergence of classical physics, where all is logically entailed. Propositional language is thus enabling, compared to metaphors, it enables logic, but at the price of categorizing, a fraught issue, the world in arbitrary ways.

But if we must live forward in time and often cannot know what can happen, we cannot reason about it with true false propositions, so again how do we do it? One way is with metaphors which are neither true nor false, and cannot be exhausted by any prestated set of true false propositions. Wittgenstein's irreducible "language games" hint in the same direction,(23). We live our lives, as he argued, in part by these irreducible language games. But more broadly and deeply is to ask: What is intuition? Is it merely savvy imagination eg of new combinations of the old, a horse with a human head? I think not. We intuit new "wholes" , much as W. James worried about new wholes in combinations of his "atoms of consciousness", and and we find "new uses of tools" all the time. My own intuition, is that intuition may be a direct awareness or participation in the Possible, Heisenberg's Potentia(2), my Res potentia, (1), that does not obey the law of the excluded middle, even as metaphors do not obey that law, even, perhaps, without Res potentia, participation in the quantum coherent aspects of our mind-body system. Because we are not conscious of quantum coherence, I will suggest that unconscious mind may be quantum coherent, hence, I hope, 'Potentia' that are ontologically real. If so, intuition arises from the unconscious mind, and only by measurement can the intuition then become conscious, (and higher brain center), perhaps a new "unity of consciousness whole" via entanglement and measurement, an hypothesis that may be testable someday, if not already in part, (17). Intuition is, then, arational, neither true nor false. As we will see below, we are almost forced to the suggestion that an unconscious quantum coherent mind carries "information" by aspects of quantum mechanics in the quantum enigma itself, to which I now turn, with my own skepticism and trepidation.

**The Quantum Enigma**

No one has solved this enigma. I try, not a physicist, below. It is of very deep importance in its own right and also because it seems to depend both upon responsible free will - at least by the physicist choosing the experiment to do - and consciousness in measurement. In turn this depends upon measurement being real which it is on most, but not all, interpretations of Quantum Mechanics. I will assume measurement is real. Doing so, in my try at a solution, may ALSO lead to a view of the emergence of a "classical enough world" via the Quantum Zeno effect, as discussed below. Thus central issues of mind-body arise here too, free will and consciousness. My hope is to find a possible unifying role for consciousness and responsible free will from electrons and fermions at measurement up to a responsible free will and non-epiphenomenal consciousness in humans. We cannot yet test if electrons really "decide", but have the hopefully decircularized Strong Free Will theorem to rest on, with the further hypothesis that the electron is conscious at measurement, now untestable.

I build on the examples in The Quantum Enigma (13).

One electron is prepared as a superposition in two boxes. If we are free willed, and thus counter factual statements can be real, we can choose to do either to an experiment to look in a box to see if the electron is there, yes or no, or instead, counterfactually, we can choose to do an experiment that allows us to "infer" that the electron is in a superposition in two boxes. Note again that the Enigma requires that counter factual statements can sometimes be true, which is ontologically possible in measurement is real and ontologically indeterminate.

On the Enigma we are, free willed, choose the question, but could have chosen otherwise, Nature answers, but could have chosen otherwise. If our consciousness plays a role in measurement, we and nature jointly "create reality" for we could have done otherwise and, if measurement is real and indeterminate, nature could have done otherwise. We do not experience this interference, we infer it. But if instead we had chosen to "look" in box 1, and we had consciously seen the electron then the electron IS in box 1. Also IF we look in first box and electron is NOT in first box where we look, it IS in the other box, ie finding electron NOT in first box "collapses the wave function" so it IS in second box, despite fact we did not look in second box. Thus, if we have free will, our choice of experiment, and here, our conscious observation of the electron, together with nature's answer, "creates" reality. Because we only "see" the electron if in box 1, our conscious experience seems to be associated with a Yes measurement answer where the wave function "collapses.

How can we use Res potentia to address the puzzle that if we look in box 1 and do not measure the electron to come to be in box 1, it comes to be in box 2? If quantum coherence of a superposition are two possibilities, then the new actual, "no electron in box 1" acausally and instantaneously alters what is now possible. In the present case, the only possibility left is that the electron is in box 2, so it comes be BE in box 2 by the new actual, "The electron is not in box 1".

Deisiderata for a solution to the Enigma:

1. What is/are the roles of consciousness in the enigma?
2. What are the roles of responsible free will and doings in the enigma?

I turn now toward what will become the new Triad: Actuals, Possibles, Mind measuring and doing in a persistent co-creative becoming.

Three central proposals to "solve the enigma", if real: 1) conscious observation is necessary and sufficient to acausally mediate measurement. This is Mind to which I add 2) Responsible doings or will, at the level of fermions exchanging bosons, and us. 3) Whatever the classical world may be, at the quantum level of that classical world, quantum variables can measure one another. So classical devices can measure via these quantum variables.

**The New Triad: Actuals, Possibles, and Mind**

I want us to consider a totally new view of quantum mechanics and reality, consisting of ontologically real actuals that obey the law of the excluded middle, ontologically real possibles that do not obey the law of the excluded middle in quantum behavior before measurement, and mind measuring and responsible free willed. In this view, measurement creates new actuals that acausally and outside of space and inside time consistent with Special Relativity, can instantaneously and acausally alter what is now possible, hence account for instantaneous changes in wave functions upon measurement and non-locality. In turn new possibles are available for mind to measure and do, creating new actuals, in a persistent cycle of quantum enigma free willed and conscious becoming in a radically participatory universe with a non-epiphenomenal "cosmic" conciousness and doings wherever measurements happen. Nothing IS, all is a persistent Becoming, a Status Nascendi.

What are we to make of these radical proposals? First, only some aspects of the above proposals have obvious merit, such as actuals acausally changing possibles instantaneously, hence altering local and non-local wave functions by altering what is now possible. We have no clear way to think about this at present, I think, without possibles. We can test if conscious observation can suffice for measurement. It will take a lot to convince us, which confront us with a major issue: No hypothesis faces the world alone, but in a vast web of hypotheses and statements of fact, chosen so that the entire web embraces our experiences of the world "at its edges" consistently. We have totally ignored the subjective pole since Newton, driven by reductionism to argue that consciousness is either an illusion (of what?) or epiphenomenal if classical physics. We are not forced to the latter, given quantum mechanics and the Poised Realm. We are not driven to no free will ontologically if measurement is real and indeterminate, for the present could have been different and I think my argument above defeat the ontological worries about quantum indeterminism and an ontologically responsible free will. I see no proof of free will. But we all assume we have it, as does the physicist who does in fact assume that he or she could, contrary to fact, have done the other Enigma experiment and we think we chose to build a rocket, could have chosen not to build it, then sent it to Mars, changed the mass of Mars, and thus the dynamics of the solar system. Did we? We all think so. If we take our subjective EXPERIENCE seriously, we can have it. We can and may ultimately, forge a new conceptual web embracing the objective and subjective pole. Who knows what further evidence we may find? And if the evolution of the biosphere precludes entailing laws, we must not think that tests must involve deduction from entailing laws. We construct the world we think we live in, metaphors, words that "mean", propositions that "mean", but do so in ways fraught with the problems Wittgenstein pointed out. We need no longer ignore our subjective pole. I am trained, in part, in medicine. No doctor fails in a neurological examination to ask the patient his or her sensations. It is time to take our subjectivity as real. I have sketched, with hope and skepticism, one route.

**Conclusions**

I have tried to show that we can only have an epiphenomenal mind if we base mind on classical physics. This is due to the causal closure of classical physics. Quantum mechanics and the Poised Realm afford two ways, decoherence and recoherence, and measurement, by which a "quantum mind" can have acausal consequences for "classical" brain and body. The Poised Realm is almost surely real and suggests an entirely new form of "computing system" a Trans Turing System which is quantum, Poised Realm and "classical". This takes us beyond the Turing machine, a finite state, finite time subset of classical physics, (1). Thus mind can be beyond the Stalemate and more than epiphenomenal, my main point. If the mind-body system is a trans-turing system it is not algorithmic, that discrete state, discrete time subset of classical physics. We are not aware of quantum superpositions but of the "yes" outcomes of measurement. I have suggested on these grounds that conscious experience is associated, testably, with measurements by molecules in the brain, whose post measurement behaviors, say in synapses, affect nearby transmembrane potentials in dendrites, hence axon firings, and then standard neurobiology; and that the shifting attention in our unity of consciousness in anatomically disconnected areas may be obtained via shifting patterns of entanglement among different sets of of quantum variables, each set addressed, for example, by widened absorption bands in decoherent quantum variables. Shifting patterns of decoherence and recoherence among quantum variables then is one way that different sets of quantum variables can be entangled. The entanglement possibility can hopefully be tied to known data from neurobiology on attention. Measurement of entangled variables is ever more highly correlated as the number of variables entangled increases, and could yield shifting Unities of Consciousness experienced as new wholes due to entanglement. None of the discussion of quantum mechanics alone leads to experiential terms. I have based my approaches to conscious experience and doing in part on the Strong Free Will theorem, and sought to remove its circularity by the non-independence of measurements of entangled quantum degrees of freedom. If this is successful, it establishes only an "ontological" basis, not yet experiential basis, for the free will of the physicist in posing the question to Nature. I then, with high skepticism, explore a new solution to the quantum enigma, assumed real, which leads to a new interpretation of quantum mechanics and reality itself: A new triad: Actuals, Possibles, and Mind as conscious observation acausally mediating measurement converting possibles to actuals that then create new possibles for mind to measure and do. While obviously highly speculative, some aspects of these ideas are testable. We are led to the possibility of panpsychism, a conscious and "responsibly deciding" universe by quantum variables upon measurement, and a "knowing" quantum coherent state which is not conscious. Perhaps this all is a version of Wheeler's Participatory Universe Observing Itself,(28).